\documentclass{aa}  
\usepackage{lscape}

\usepackage[switch]{lineno} 
\usepackage{graphicx}
\usepackage{hyperref}
\usepackage{placeins}

\usepackage{txfonts}
\usepackage{color}
\usepackage{caption}
\usepackage{multicol}
\usepackage{multirow}

\usepackage{amsmath}

\begin{document}

   \title{Molecular line emission from 1000 au scales outflows to <30 au compact structures in NGC1333 IRAS4A2.}

   \author{Osmar ~M. Guerra-Alvarado,
          \inst{1}
           N. van der Marel,
           \inst{1}
           P. Nazari,
           \inst{2}
            J. Di Francesco,
           \inst{3}
           Ł. Tychoniec,
           \inst{1}
           L.~W. Looney, 
           \inst{4}
           E.~G. Cox,
           \inst{5,6}
           D.~J. Wilner
           \inst{7},
           M. R. Hogerheijde
           \inst{1,8}
}

   \institute{Leiden Observatory, Leiden University, PO Box 9513, 2300 RA Leiden, The Netherlands
        \\      \email{guerra@strw.leidenuniv.nl}
    \and European Southern Observatory, Karl-Schwarzschild-Strasse 2, 85748 Garching, Germany
    \and      National Research Council of Canada, Herzberg Astronomy and Astrophysics Research Centre, 5071 West Saanich Road, Victoria, BC V9E 2E7, Canada
    \and          Department of Astronomy, University of Illinois, 1002 West Green Street, Urbana, IL 61801, USA
    \and          Center for Interdisciplinary Exploration and Research in Astronomy, Northwestern University, 1800 Sherman Rd., Evanston, IL, 60202, USA
    \and          NSF MPS-Fellow
    \and         Center for Astrophysics | Harvard \& Smithsonian, Cambridge, MA 02138, USA
    \and Anton Pannekoek Institute for Astronomy, University of Amsterdam, PO Box 94249, 1090 GE, Amsterdam, The Netherlands
}

  \abstract
   {}
   {Studying protostellar objects in their earliest stages, particularly during the Class 0 phase, provides key insight into the beginnings of planet formation and dust evolution. Disentangling the various components, such as the envelope, outflow, and nascent disk, to characterize and understand these young systems, however, is particularly challenging. High spatial and spectral resolution observations of molecular line emission with the Atacama Large Millimeter/submillimeter Array (ALMA) are therefore crucial for probing their complex environments.}
  {In this work, we present high-resolution ($\sim$ 30 au) ALMA observations at 1.3 millimeters of the Class 0 protostellar system IRAS4A2. We analyze the gas emission surrounding this young source, tracing it from the extended outflow to the most compact inner region and identifying emission lines using the spectral analysis tool CASSIS.} 
  {We detected large, well-traced outflows in $\rm HCN~(3-2)$, $\rm H_{2}CO~(2_{1 ~2}-1_{1~1})$, and $\rm HCO^{+}~(3-2)$, along with numerous complex organic molecules (COMs) such as $\rm C_{2}H_{3}CN$, $\rm CH_{2}(OH)CHO$, $\rm CH_{3}OCHO$, and $\rm CH_{3}C^{15}N$ tracing central, more compact regions. These molecules span upper-energy levels (E$_u$) ranging from 20 to 487 K and have excitation temperatures between 100 and 300 K. Using moment maps, we analyzed the kinematics and spatial distributions of the molecular emission, revealing a wide range of spatial scales, from compact structures within the IRAS4A2 core at $\sim$8 au in radius, to extended $\sim$5,000 au outflow emission. Specifically, we find that $\rm CH_{3}CDO$ and $\rm CH_{3}OCHO$ could be both good tracers of the disk, possibly tracing its rotation. Lines of $\rm OCS~(22-21)$, $\rm SO_2~(13_{3~11}-13_{2~12})$, $\rm HCN$, $\rm H_2CO$, and $\rm HCO^{+}$ with lower upper-level energies, show more extended structures around IRAS4A2, likely tracing the envelope, disk, accretion shocks, the base of an outflow, and the outflow itself. Some of these molecular lines exhibit signatures consistent with Keplerian rotation, indicating a central protostellar mass of approximately 0.2 M$_{\odot}$.}
  {Most COMs appear to trace distinct inner regions near the central protostar, while other molecules like $\rm OCS$, $\rm SO_{2}$, $\rm HCN$, and $\rm H_2CO$ trace more extended structures, such as the envelope or outflows. The kinematics, emission patterns, and position–velocity diagrams suggest that individual molecules trace multiple components simultaneously, making it challenging to disentangle their true origins. Altogether, these findings highlight the complex spatial distribution within the IRAS4A2 system.} 

   \keywords{Planetary systems: Protoplanetary disk - Radio lines: planetary systems
               }

\titlerunning{Molecular line emission in NGC1333 IRAS4A2.}
\authorrunning{O. Guerra-Alvarado et. al.}
   \maketitle

\section{Introduction}

Studying and understanding the evolution of protoplanetary disks from very embedded Class 0 objects to Class II disks is fundamental to unraveling the processes of planet formation. Protoplanetary disks emerge at very early evolutionary stages \citep{2009A&A...507..861J,2014A&A...562A..77H,2019A&A...621A..76M,2020ApJ...890..130T}, with evidence that suggests that planet formation may already be underway during these initial phases \citep{2018NatAs...2..646H, 2020A&A...640A..19T}. Although many of these early-stage disks appear smooth in current observations \citep{2024ApJ...973..138H}, substructures such as rings, gaps, and spirals could already be forming but remain undetected or hidden due to resolution limitations, optical depth effects \citep{2023ApJ...951....8O} and temperature effects \citep{2025A&A...697A..84N}. Even though studying planet formation in Class 0/I sources remains a challenge, primarily because of the optically thick dust continuum emission, it is still possible to study the chemical composition of these young disks, which can provide crucial insights into their evolution.

In this context, numerous Class 0/I protostars have been identified with hot, dense environments known as hot corinos, which are analogs of the hot cores found in high-mass star-forming regions \citep{2004ASPC..323..195C}. Many of these sources have been found to contain high abundances of Complex Organic Molecules (COMs), species with six or more atoms \citep{2009ARA&A..47..427H, 2012A&ARv..20...56C, 2023ASSP...59....3C}. Most of these studies, however, have been conducted at low spatial resolution, limiting our ability to resolve the detailed structure and stratification of molecular species within these young disks. 
\begin{table*}[ht]
\centering
\caption{Data cubes characteristics}
\label{tab:datacubes}
\begin{tabular}{lcccc}
\hline\hline
 Observations & Rest Frequency (GHz) & Beam Size ($\arcsec$) & Channel Width (km\,s$^{-1}$) & RMS (mJy\,beam$^{-1}$) \\
\hline
O6B1 short baseline&267.557 & $0.28 \times 0.17$ & 0.5 & 2.8 \\
O6B2 short baseline& 265.886 & $0.28 \times 0.17$  & 0.5 & 2.87 \\
Band 4 short baseline&140.839 & $0.44 \times 0.28$ & 1 & 2.3 \\
\hline
\hline

O6B1 long+short baseline&267.557 & $0.13 \times 0.085$ & 0.5 & 1.59 \\
O6B2 long+short baseline& 265.886 & $0.155 \times 0.10$  & 0.5 & 1.57 \\
Band 4 long+short baseline&140.839 & $0.10 \times 0.047$ & 1 & 0.95 \\

\hline
\end{tabular}
\end{table*}
With the unprecedented sensitivity and resolution of the Atacama Large Millimeter/submillimeter Array (ALMA) we are now able to investigate in more detail the molecular distribution of these hot corinos. 

As of today, only a handful of Class 0 hot corino sources have been resolved at high angular resolution ($\sim$ 0.1") (IRAS 16293-2422 A \citep{2022ApJ...941L..23M,2020ApJ...897...59M}, SVS13-A \citep{2022ApJ...928L...3B}, IRAS2AA \citep{2022Natur.606..272J}, B335 \citep{2022ApJ...935..136O,2022ApJ...930...91D, 2025A&A...696A...1H}, IRAS4A2 \citep{2025A&A...695A..78F}, HH212 \citep{2022ApJ...937...10L}) and L1448-mm \citep{2024A&A...686A.201N}. Interestingly, in HH 212,  \citet {2022ApJ...937...10L} revealed molecular stratification, which was attributed to the binding energies of the molecules (sublimation temperature), as the temperature is expected to increase closer to the host star. Studying these young protoplanetary disks, where molecules sublimate at specific temperatures in different regions, can provide insights into where planet formation and dust growth may already be occurring, which we aim to identify in IRAS4A2.

NGC1333 IRAS 4A2 was identified as the second hot corino source ever by \citet{2004ApJ...615..354B}. It is located at a distance of 293 pc \citep{2018ApJ...869...83Z} in the NGC1333 region in Perseus, and it is part of a binary Class 0 system, separated by $1{\farcs}8$ from its primary component, IRAS4A1 \citep{2018ApJ...867...43T}. IRAS4A2 has been extensively studied in the literature \citep{2010ApJ...723L..34C, 2015ApJ...804...81T, 2017A&A...606A.121L, 2017A&A...599A.121D, 2021ApJ...916...82C, 2024A&A...686L..13D, 2024MNRAS.534L..48C}, revealing a rich inventory of complex organic molecules (COMs) and strong absorption features that partially affect its emission \citep{2020A&A...640A..75D}. Both IRAS4A1 and IRAS4A2 share a common envelope with a total mass of approximately 8 M$_\odot$ \citep{2019A&A...621A..76M}, which contributes significantly to the millimeter continuum emission. The inner regions of IRAS4A2 are believed to be optically thick at millimeter wavelengths \citep{2019A&A...632A...5G, 2020ApJ...889..172K}. The system hosts well-characterized bipolar outflows observed in various tracers, including CO, SiO, and SO \citep{2011PASJ...63.1281C, 2015A&A...584A.126S, 2016ApJ...819..159C, 2021ApJ...916...82C}. An S-shaped morphology has been observed in the IRAS4A2 outflow, along with an inverted velocity gradient in the inner regions compared to the large-scale outflow, as traced by SO \citep{2015A&A...584A.126S, 2021ApJ...916...82C}. This inner velocity gradient is consistent with that traced by ammonia emission \citep{2010ApJ...723L..34C}. Altogether, these features underscore the complexity and rich structure of this particular deeply embedded Class 0 protostellar system.

In this work, we present a detailed study of the IRAS4A2 hot-corino, from its large-scale outflows to its innermost compact emission. In Section~\ref{sec:Observations}, we describe the observations and the process to obtain the spectral line emission. Section~\ref{sec:Results} discusses our line identification method, along with the techniques used to create the images of the outflows and the extended and compact emission. Finally, in Section~\ref{sec:Discussion}, we explore the possible spatial distribution of these molecules and their implications for IRAS4A2 in the context of dust, disk evolution, and planet formation.

\section{Observations}\label{sec:Observations}

The spatially resolved line observations for IRAS4A2 were obtained using the Atacama Large Millimeter/submillimeter Array (ALMA). Specifically, Band 4 and Band 6 data were gathered under the project code 2018.1.00510.S (PI: James Di Francesco). For a more comprehensive understanding of the observations, including details about data characteristics, calibration, self-calibration, and imaging of the continuum emission, refer to \citet{2024A&A...681A..82G}

Line spectral windows were separated from the continuum ones, and in this instance, no channel averaging was applied. The self-calibration tables were derived from the continuum analysis and subsequently used in the line spectral windows. In Band 6, observations were carried out using two spectral settings, which we refer to as O6B1 and O6B2 for simplicity. O6B1 has a rest frequency of 267.557 GHz and encompasses two spectral windows (SPWs), each featuring 960 channels. These SPWs have central frequencies of 267.568 GHz and 267.569 GHz, respectively. O6B2, on the other hand, has a rest frequency of 265.886 GHz with four SPWs, each with 960 channels and central frequencies of 265.874 GHz, 265.894 GHz, 265.896 GHz, and 265.898 GHz, respectively. In total, the Band 6 data of O6B1 and O6B2 cover frequency ranges from 265.8214 GHz to 265.9385 GHz and from 267.4925 GHz to 267.6103 GHz, respectively.

The Band 4 line data have a rest frequency of 140.839 GHz with two SPWs, each containing 1920 channels with a central frequency of 140.795 GHz and a frequency range between 140.7319 - 140.8512 GHz. Across all observations, the system velocity was assumed to be 6.95 km/s \citep{2001ApJ...562..770D}, which was later used for the analysis.
 
The final data cubes were produced using the \textit{tclean} task in CASA, with \textit{specmode} set to "cube" and the deconvolver to "Multiscale." We used scale values of 0, 2.5, 10, and 20 times the pixel size, which was fixed at 0.02" for all line cubes, approximately ten times smaller than the beam size. The spectral resolution is 0.5 km/s for Band 6 observations and 1 km/s for Band 4. These values were selected to balance sensitivity with maintaining the highest possible spectral resolution. Briggs weighting with a robust parameter of 0.5 was adopted for both datasets, as it was found to be the optimal choice based on the cubes images and also in the continuum images presented in \citet{2024A&A...681A..82G}. It also provided the best resolution against other robust values for determining the spatial distribution of the molecular emission, without blurring or missing any large-scale structures or weaker lines. This approach was confirmed through comparison with images made via natural weighting. The final characteristics are shown in Table~\ref{tab:datacubes} with typical beam sizes of 0.1" and rms values of $\sim$1 mJy/beam per 1 km/s channel.

Finally, for the short baseline data, continuum subtraction was performed prior to imaging using the \textit{uvcontsub} task in CASA, resulting in three short baseline cubes. For the concatenated data, which has higher resolution and greater sensitivity to small features, determining the continuum level was more challenging. To enhance the accuracy of continuum subtraction and better determine the continuum emission in the image cubes, the STATCONT tool \citealp{2018A&A...609A.101S} was employed. STATCONT utilizes a 'corrected sigma-clipping algorithm' for determining continuum emission in line-rich sources like IRAS4A2. This method was applied to all concatenated cubes in the study, resulting in the creation of three other cubes. We used only the short-baseline data to account for the most extended emission, to study the large-scale outflows, and the concatenated (short plus large baseline) data to study the emission surrounding IRAS4A2.

\section{Results}\label{sec:Results}

Our data revealed emission lines from $\rm SO_{2}~13_{3 ~11}-13 _{2~ 12}$, $\rm OCS~22-21$, $\rm HCN~3-2$, $\rm H_{2}CO~2_{1 ~2}-1_{1~1}$ and $\rm HCO^{+}~3-2$, along with several COMs with upper energy levels of 20 to 487 K (see Table~\ref{tab:IRAS4A2 Molecules First Table}). These emission lines exhibit a variety of spatial components and kinematic structures. In the following section, we first describe the identification of the emission lines around IRAS4A2, we then characterize the spatial distribution of the integrated emission and analyze the kinematics, beginning with the outflow, followed by the extended emission, and concluding with the compact COM emission.

\subsection{Analysis and line detection with CASSIS}

We conducted an exploration of the spectral emission windows within the concatenated cubes and identified several emission lines originating from a small region near the center of IRAS4A2. Across our cubes, we observed some strong absorption features towards a small region inside the continuum emission. To mitigate this, spectra from all of our data were extracted within CARTA (Cube Analysis and Rendering Tool for Astronomy) \citep{angus_comrie_2018_3377984} by selecting a pixel as close as possible to the continuum peak ($\sim$ 0.08", 23 au separation), but still outside the region affected by absorption, located at Dec = 31:13:31.9791986 and RA = 3:29:10.4336922. Subsequently, these spectra were converted into brightness temperature and processed for analysis using the CASSIS spectral analysis tool version 5.1.1 \citep{2015sf2a.conf..313V}. Following the methodology outlined in \citet{2022A&A...668A.109N}, we assumed local thermodynamic equilibrium (LTE) conditions and employed a fitting-by-eye approach. Potential transitions were investigated within CASSIS, utilizing the Cologne Database for Molecular Spectroscopy (CDMS, \citet{2001A&A...370L..49M}) and the Jet Propulsion Laboratory (JPL) database \citep{1998JQSRT..60..883P}.

Commonly detected molecules and COMs were prioritized in the search, with upper energy state levels and Einstein coefficient limits set to E$_{up} \leq 800$ K and A$_{ul} \geq$ 9x$10^{-6}$ s$^{-1}$. Since we extracted the spectra from single pixels, the source size parameter in CASSIS was uniformly set to 1000 for all lines, i.e., unity beam filling factors were assumed. The Full Width at Half Maximum (FWHM) of each line was adjusted to fit the spectra better, and the V$_{lsrk}$ was uniformly set to 6.95 km/s for all spectral cubes.

\begin{figure*}[ht!]

\sidecaption
\includegraphics[trim=0 3cm 1cm 0, clip=true,width=12cm]{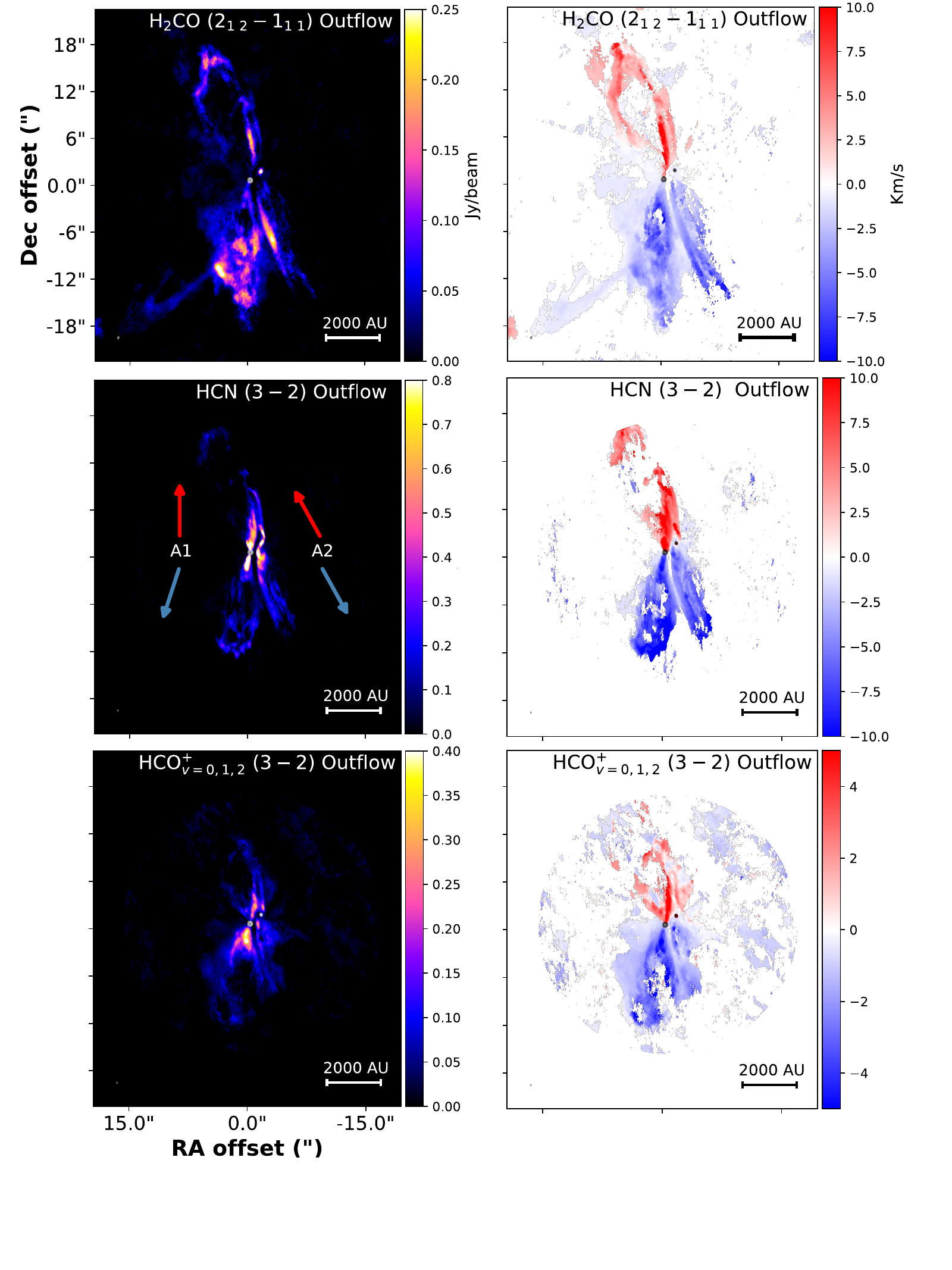}
\caption{Integrated emission and velocity maps of the three main outflow tracers in our spectra: $\rm HCN$, $\rm H_{2}CO$, and $\rm HCO^{+}$ generated using the short baseline line data for IRAS4A. The emission from the outflows of both sources, IRAS4A1 and IRAS4A2, is resolved and can be distinguished separately. In the figures, the dust continuum emission of IRAS4A1 and IRAS4A2 is outlined in small white contours, showing 5 $\sigma$ and 10 $\sigma$. The beam is drawn in the lower left part of each panel, 0.28"$\times$0.17" for $\rm HCN$ and $\rm HCO^{+}$ and 0.44"$\times$0.28" for $\rm H_{2}CO$. The red and blue arrows in the middle-left panel indicate the outflow directions of IRAS4A1 (left) and IRAS4A2 (right)\citep{2024MNRAS.531.2653C}, respectively.} \label{fig: Outflows}
\end{figure*}

Following the criteria from \citep{1998JQSRT..60..883P}, a molecule was considered detected if at least three lines were found at 3-$\sigma$; otherwise, it was categorized as a `tentative detection'. Figure~\ref{fig: Cassis Model} displays the observed spectra, where lines were identified, from our cubes in IRAS4A2 alongside our CASSIS model of the lines. Additionally, Appendix~\ref{appendix:a} includes Table~\ref{tab:IRAS4A2 Molecules First Table}, which provides detailed information on all identified molecular lines and their transitions, and Table~\ref{tab:IRAS4A2 Molecules Second Table}, which summarizes their extent and physical properties, including excitation temperatures and column densities.

Many COMs show little to no absorption toward IRAS4A2, whereas in IRAS4A1 their emission appears completely absorbed by the dust in our data. This difference may reflect their evolutionary stages, with IRAS4A1 being younger ($< 10^3$ yr; \citealt{2015A&A...584A.126S}) and IRAS4A2 older ($> 10^4$ yr; \citealt{2010ApJ...723L..34C}). Recent studies, such as \citet{2020ApJ...896L...3D}, \citet{2017A&A...606A.121L}, and \citet{2024A&A...681A..82G}, have highlighted the differences in optical depth between the two sources and even some COM emission toward IRAS4A1 has been observed in absorption against a hot, optically thick dust background \citet{2019ApJ...872..196S}, which may explain the lack of emission in our data of A1. Despite the challenge in observing emission from IRAS4A1, analyzing the diverse species emanating from IRAS4A2 offers valuable insights into the early stages of Class 0 objects and the beginnings of protoplanetary disk formation. 

\begin{figure*}[ht!]

\sidecaption
\includegraphics[trim=0 0cm 0cm 0, clip=true,width=12cm]{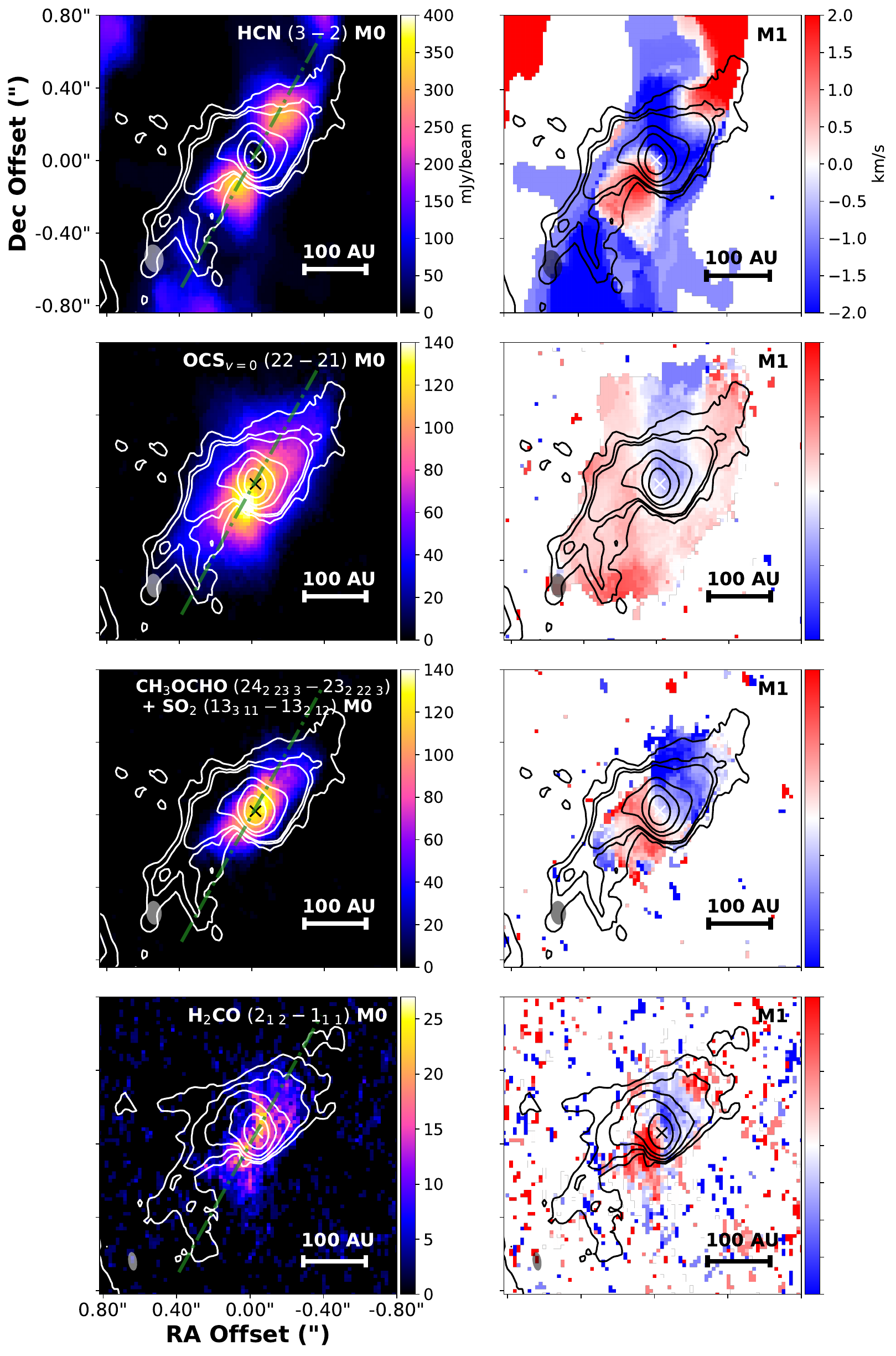}
\caption{Integrated emission and velocity maps of $\rm SO_{2}$, $\rm OCS$, $\rm H_{2}CO$ and $\rm HCN$ using the long+short baseline data. These molecular lines trace a larger extended emission in IRAS4A2. We identified these features visually from the spectra as the most intense and spatially extended emission lines. The white and black contours represent the 6 to 60-$\sigma$ levels of the continuum emission from the ALMA Band 6 and Band 4 data, highlighting the underlying dust distribution. The synthesized beam is shown in the lower-left corner of all panels, and the green dashed lines indicate the cuts used to extract the position–velocity (P–V) diagrams perpendicular to the outflow axis.}
\label{fig: ExtendedEM}
\end{figure*}

\begin{figure*}[t!]
\centering
\includegraphics[trim=0 0cm 0cm 0cm, clip=true,width=1.01\textwidth]{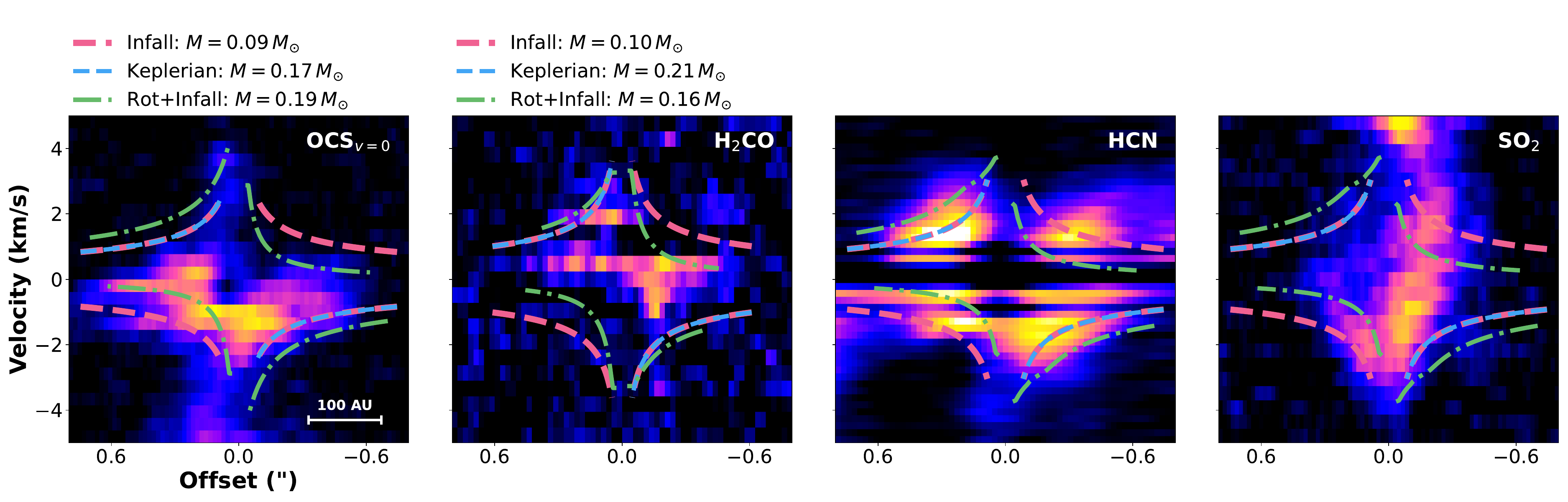}
\caption{Position-velocity diagrams of $\rm SO_{2}$, $\rm OCS$, $\rm H_{2}CO$, and $\rm HCN$ extracted along a cut perpendicular to the jet axis. For $\rm{OCS}$ and $\rm{H_2CO}$, we fitted the emission using three kinematic models: pink lines represent the infall motion fit, blue lines correspond to the Keplerian rotation fit, and green lines indicate the rotating infalling envelope model. The labels show the mass of the central protostar derived from each fit, and the negative offset marks the north-west direction. For $\rm{SO_2}$ and $\rm{HCN}$, we overlay the model with the average of the fit parameters obtained from the models in $\rm{OCS}$ and $\rm{H_2CO}$ for comparison, as their emission is more complicated. } \label{fig:PV-diagramas}
\end{figure*}

Our IRAS4A2 CASSIS model confirmed the detection of four COMs: $\rm C_{2}H_{3}CN$ (Vinyl cyanide), $\rm CH_{2}(OH)CHO$ (Glycolaldehyde ), $\rm CH_{3}OCHO$ (Methyl formate), and $\rm CH_{3}C^{15}N$ (Methyl cyanide). While we were unable to fully fit some of the emission around IRAS4A2, nearly all lines were covered within our frequency range. Tentative detections were also made: the $\rm aGa'$ glycol line, while providing temperature and column density estimates, was consistently blended with other molecules. For $\rm C_{2}H_{5}^{13}CN$, only one detection was made, but two tentative features are consistent only with warmer and denser gas than included in our fit. Finally, the higher frequency line of $\rm CH_{3}NCO$ is partially cut off in the frequency range, leading to underestimations of the associated excitation temperature and column densities. Despite comparable column densities with those found in \citet{2017A&A...606A.121L} for different COMs, it has already been mentioned before that the high dust opacity in hot corinos results in underestimations of the abundances \citep{2020ApJ...896L...3D}.
Regardless of these limitations, this work presents the first detection of $\rm C_{2}H_{3}CN$ and $\rm aGg'$-glycol around IRAS4A2.

\subsection{Outflows, extended emission and compact emission in IRAS4A2}

\subsubsection{Outflows}

We employed the code \textit{bettermoments} for collapsing the concatenated spectral cube, generating moment maps of the outflows of the IRAS4A system (both components) and the emission lines around only IRAS4A2 \citep{2018RNAAS...2c.173T}.
For a comprehensive view of the large-scale outflows, moment 0 and moment 1 images were generated for $\rm HCN$, $\rm HCO^{+}$, and $\rm H_{2}CO$ using the short baseline data. We present these in Figure~\ref{fig: Outflows}, providing information about the spatial distribution and velocity structures of these outflows. 

These outflows extend to approximately 18" ($\sim$5300 AU). We resolve both outflows in all lines which allows for a clear distinction between the emission that originates from IRAS4A1 and IRAS4A2. We observed the distinctive S-shaped (observed in the moment 0 maps) pattern of the IRAS4A2 outflow, consistent with previous findings by \citet{2015A&A...584A.126S} and \citet{2021ApJ...916...82C}. This shape is attributed to a misalignment between the initial core angular momentum vector and the magnetic field. It is suggested that this misalignment can happen if the Hall-induced magnetic field flips the angular momentum vector between the flattened envelope and the circumstellar envelope \citep{2020MNRAS.492.3375Z,2016MNRAS.457.1037W}.

\subsubsection{Extended emission on $\sim$ 100 au scales}

Four lines were identified as tracers of more extended emission in the concatenated data, $\rm SO_{2}$, $\rm HCN$, $\rm H_{2}CO$, and $\rm OCS $, each exhibiting its own distinct morphology. In Figure~\ref{fig: ExtendedEM} we present the moment 0 and 1 maps of these molecules, with white contours outlining the continuum dust emission at 1.3 mm. All molecular emission images of IRAS4A2 are centered at RA = 3:29:10.43, Dec = 1:13:32.01. For these specific spectral lines, we obtained position-velocity (P-V) diagrams to determine the nature of the observed emission, whether it was driven by infall motion, Keplerian rotation from a disk, or a rotating infalling flattened envelope. To analyze this, we used CARTA. We extracted the emission from a cut perpendicular to the outflow axis of IRAS4A2 (29°, \citealt{2024MNRAS.531.2653C}), extending from the North-East to the South-West. We analyzed the $\rm H_{2}CO$, and $\rm OCS$ image by selecting 24 points along the edge of the emission by eye and fitted them using a non-linear least squares fitting procedure and the following models:

\noindent
For infall motion, we adopt
\begin{equation}
v_{\text{infall}}(r) = \sqrt{2} \cdot \sqrt{\frac{G M}{|r|}},
\end{equation}
For Keplerian rotation, we use
\begin{equation}
v_{\text{Keplerian}}(r) = \sqrt{\frac{G M}{|r|}},
\end{equation}
and for a rotating infalling flattened envelope, following the methodology in \citet{2014Natur.507...78S} and \citet{2017ApJ...843...27L}, we adopt
\begin{equation}
v_r(r) = -\sqrt{\frac{2GM}{r} - \frac{l^2}{r^2}},
\end{equation}
and,
\begin{equation}
v_\phi(r) = \frac{l}{r},
\end{equation}
where $G$ is the gravitational constant, $M$ is the mass of the central protostar, $l$ is the specific angular momentum, and $r$ is the distance from the central protostar. For the infall and Keplerian models, the central protostellar mass was treated as a free parameter. In the rotating infall model, both the protostellar mass and the specific angular momentum $l$ were fitted.

For $\rm{SO_2}$ and $\rm{HCN}$, the emission is more complex than what our simple models can fully explain. Therefore, we use the average of the fits from $\rm{H_2CO}$ and $\rm{OCS}$ to examine whether any regions of these molecules' emission remain consistent with these models.

In Figure~\ref{fig:PV-diagramas}, we present the P–V diagrams for the four extended molecules, overlaid with Keplerian, infall, and rotating infalling envelope models. While $\mathrm{H_2CO}$ and $\mathrm{OCS}$ tend to yield relatively low protostellar mass estimates, their values are generally consistent with each other (see mass labels in Figure~\ref{fig:PV-diagramas}). They also show similar angular momentum ($l$) and centrifugal barrier (CB) radii, which are summarized in Table~\ref{tab:dyn_properties}. These similarities provide encouraging evidence that $\rm{H_2CO}$ and $\rm{OCS}$ may originate from the same physical component, given their comparable dynamical structures.

\begin{table}[hb]
\centering
\small 
\caption{Derived properties for each molecule under different dynamical models. RIE stands for rotating infalling envelope.}
\begin{tabular}{l|l|c|c|c}
\hline
Molecule & Model & M [$M_\odot$] & $l$ [au$\cdot$km$\cdot$s$^{-1}$] & CB [au] \\
\hline
\multirow{3}{*}{OCS} 
    & Keplerian & 0.17$\pm$0.07& -- & -- \\
    & Infall & 0.09$\pm$0.03 & -- & -- \\
    & RIE & 0.19$\pm$0.04 & 38.5$\pm$7.8 & 4.3$\pm$2 \\
\hline
\multirow{3}{*}{H$_2$CO} 
    & Keplerian & 0.21$\pm$0.04 & -- & -- \\
    & Infall & 0.1$\pm$0.02 & -- & -- \\
    & RIE & 0.16$\pm$0.03 & 44.4$\pm$7.74 & 6.8$\pm$2.7 \\
\hline
\end{tabular}
\label{tab:dyn_properties}
\end{table}

There are, however, notable differences in the best-fit models for $\rm OCS$ and $\rm H_{2}CO$, as well as in the overlay comparisons for $\rm HCN$ and $\rm SO_{2}$, which may indicate distinct kinematic structures or physical origins. For $\rm OCS$ and $\rm H_{2}CO$, the rotating infalling flattened envelope model reproduces the emission reasonably well. However, in the case of $\rm H_{2}CO$, the model fails to capture the low-velocity emission, which could suggest that a purely Keplerian model may be more appropriate. In contrast, for $\rm HCN$ and $\rm SO_{2}$, both the Keplerian and pure infall models reproduce certain features of the observed emission patterns. Nevertheless, significant emission that is not well described by any of the models considered remain. Some possible contributing factors here are: absorption in certain velocity channels, that the molecules could be tracing multiple kinematic components, or that the molecular lines fitted are blended with other molecules, particularly for $\rm SO_{2}$, which is blended with $\rm CH_{3}OCHO$. Contamination from $\rm CH_{3}OCHO$ in the $\rm SO_{2}$ line is difficult to quantify, as it appears to be completely blended in velocity. The expected peak of $\rm CH_{3}OCHO$ emission lies near 3 km/s in the $\rm SO_{2}$ P-V diagram, however it is located so close to the $\rm SO_{2}$ and $\rm OCS$ features that it becomes challenging to distinguish where one line ends and the other begins, let alone to accurately assess the contribution from each component.

Notably, the emission that fits between the chosen models aligns well with previous observations of $\rm NH_{3}$ \citep{2010ApJ...723L..34C}. In contrast, the emission in other quadrants in $\rm HCN$ and $\rm SO_{2}$ that cannot be reproduced by any of the models shows a distinct change in the velocity gradient. This shift coincides with the outflow position, as seen in the moment 1 maps in Figure~\ref{fig: ExtendedEM}. Similar structures have been reported in other tracers associated with outflows, including $\rm SiO$ \citep{2024A&A...686L..13D}, $\rm H_{2}CO$ \citep{2019ApJ...885...98S}, and $\rm CO$ and $\rm SO$ \citep{2021ApJ...916...82C}, suggesting that these molecular lines are likely tracing multiple dynamical components.

Finally, by examining the moment 1 maps, we find that the velocity fields in the inner regions of $\rm HCN$, $\rm H_{2}CO$, $\rm OCS$, and $\rm SO_{2}$ are misaligned (flipped) with those of the outflow (see Fig.~\ref{fig: ExtendedEM}). This inversion of the velocity maps has been previously observed in various disks before, such as VLA 1623B \citep{2022ApJ...927...54O}, BHR71 \citep{2019ApJ...870...81T}, and IRAS 16293-2422 \citep{2013ApJ...764L..14Z}. Multiple origins for this inversion have been proposed, including interaction with the outflow, hydrodynamical interaction with the circumbinary disk, accretion shocks, and, more recently, material accelerated by the disk wind as it crosses the plane of the sky \citep{2024A&A...686L..13D}. The exact cause of the inverted velocity gradient, however, remains incompletely understood and requires further investigation.

\begin{figure*}[t]
\centering

\includegraphics[trim=0 0cm 0cm 0cm, clip=true,width=\textwidth]{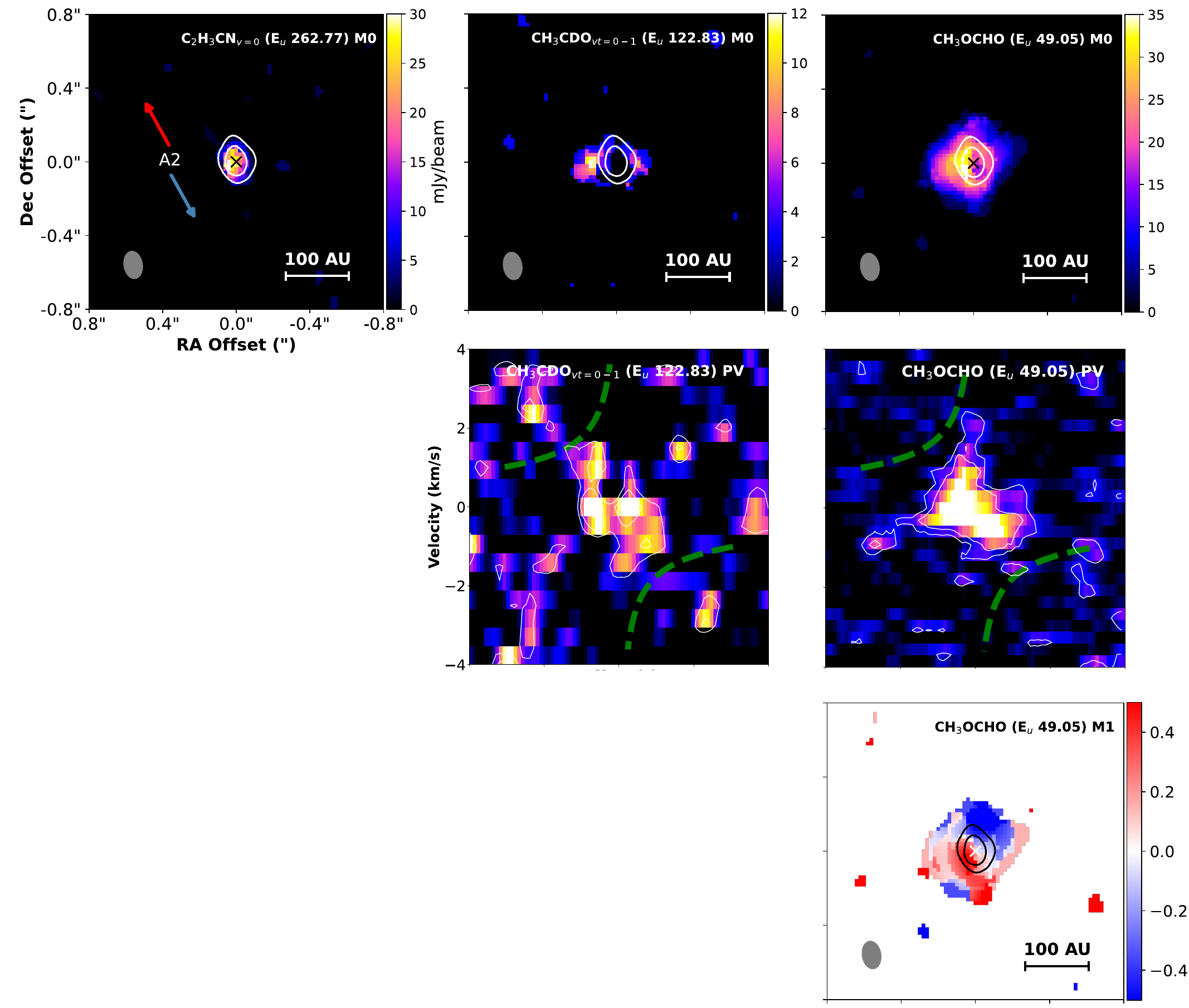}
\caption{Integrated intensities of the three representative molecular lines that were not blended with any other, showcasing the three visually identified regions they trace within the IRAS4A2 disk: $\rm C_{2}H_{3}CN$, $\rm CH_{3}CDO$ and $\rm CH_{3}OCHO$. The white contours show the 30 $\sigma$ to 60 $\sigma$ levels of the continuum emission, and the synthesized beams are indicated in the lower left corner of each image. The middle panels  shows a P–V diagram of $\rm CH_{3}CDO$ and $\rm CH_{3}OCHO$, extracted along a cut perpendicular to the outflow axis. The green dashed curve represents the Keplerian rotation model from $\rm H_{2}CO$, while white contours indicate emission levels from 1.5 $\sigma$ to 3 $\sigma$, based on the Band 6 line data. The bottom panel shows the moment 1 (M1) map of $\rm CH_{3}OCHO$, indicating the velocity structure across the emitting region.} \label{fig: CompactEM}
\end{figure*}

\subsubsection{Compact emission on sub-100 au scales}

Several COMs, like $\rm CH_3OCHO$, $\rm CH_{3}CDO$, $\rm C_{2}H_{3}CN$,$\rm CH_3NCO$, $\rm CH_3{}^{18}OH$, and more, show much more compact emission, on scales of 20 to 50 au with a range of morphologies (see Fig~\ref{fig: Moment0maps}). 
As an illustration, Figure~\ref{fig: CompactEM} displays the moment 0 maps generated by \textit{bettermoments} for three representative molecules that were not blended with any other line emission. These maps showcase three different morphologies traced within the IRAS4A2 system, indicating that the compact COM emission may arise from different regions like the disk, a warm inner envelope, or both \citep{2021A&A...655A..65T}. However, with the current data, a definitive distinction cannot be made. In addition, Figure~\ref{fig: CompactEM} includes the P-V diagram of $\rm CH_{3}CDO$ and $\rm CH_{3}OCHO$, taken along a cut perpendicular to the outflow axis, and the moment 1 map of $\rm CH_{3}OCHO$ showing the velocity structure of this particular molecule.

To investigate the spatial distribution of molecular emission in IRAS4A2, we analyzed the emission radius using aperture photometry on the moment 0 maps. We applied a signal-to-noise ratio (SNR > 5) threshold, masking all values below this limit, and then determined the 68\% flux radius of the emission, r$_{68}$. We applied the methodology described in Appendix D of \citet{2021A&A...648A..45P}, where we extract a cut perpendicular to the outflow axis going through the continuum peak to measure the spatial extent of the emission. We then assume the measured r$_{68}$ corresponds to the FWHM/2 of the emission profile and deconvolve this value from the beam size to correct for beam smearing and avoid biases due to intensity variations across different molecular lines. Figure~\ref{fig: TempsAndDistribution} shows the resulting radius plotted against the excitation temperature derived from CASSIS for molecules with available data. Using the same approach, Figure~\ref{fig: TempsAndDistribution} also shows the measured radius of each molecule plotted against its respective upper energy levels.

It is important to note that many observed emission lines are blended with others, complicating the estimation of their emission radii and reducing accuracy. Furthermore, some molecular emission exhibits low signal-to-noise ratios or remains unresolved at the current spatial resolution. In these cases, we fixed the FWHM to the minimum beam size to avoid overestimating their spatial extent. Despite these concerns, we chose to proceed with the analysis, as even blended lines, when observed at high resolution, can still provide valuable insights into the chemical structure of young Class 0 objects like IRAS4A2. For a more precise determination of temperature and radius, multiple transitions of unblended molecular emission lines are required, as demonstrated in \citet{2025A&A...695A..78F}. 

We note from Figure~\ref{fig: TempsAndDistribution} that the estimated temperatures are broadly consistent with those reported by \citet{2025A&A...695A..78F}. However, one peculiar result arises for $\rm C_2H_5{}^{13}CN$ (Ethyl cyanide, propionitrile). Given its chemical similarity, we would expect this molecule to exhibit comparable behavior to $\rm C_2H_3CN$ (Vinyl cyanide). While both molecules display similar emitting radii, their derived excitation temperatures differ significantly. In the case of $\rm C_2H_5{}^{13}CN$, two of the transitions used in the analysis are blended. It is therefore possible that we underestimated the intensity contribution of these blended transitions, which may have resulted in an underestimation of the excitation temperature. Furthermore, we acknowledge that $\rm C_2H_5{}^{13}CN$ is an isotopologue, unlike $\rm C_2H_3CN$, which could also contribute to the observed differences and should be considered in the interpretation.

\section{Discussion}\label{sec:Discussion}

\subsection{Spatial scales of the dust and gas of IRAS4A2}

In IRAS4A2, the distributions of emission from COMs and other molecules appear to trace distinct physical processes. Identifying these at high resolution can reveal key regions where planet-forming and disk-forming processes occur, as well as the sublimation temperatures that dictate where certain molecules transition from solid to gas. In particular, around IRAS4A2, molecular emission extends from the most inner regions ($\sim$ 10 au) to large-scale outflows ($\sim$ 5300 au).

\subsubsection{Extended emission and its structure in IRAS4A2}
\begin{figure*}[ht!]
\centering

\includegraphics[trim=0.2cm 0cm 0.0cm 0.0cm, clip=true,width=1\textwidth]{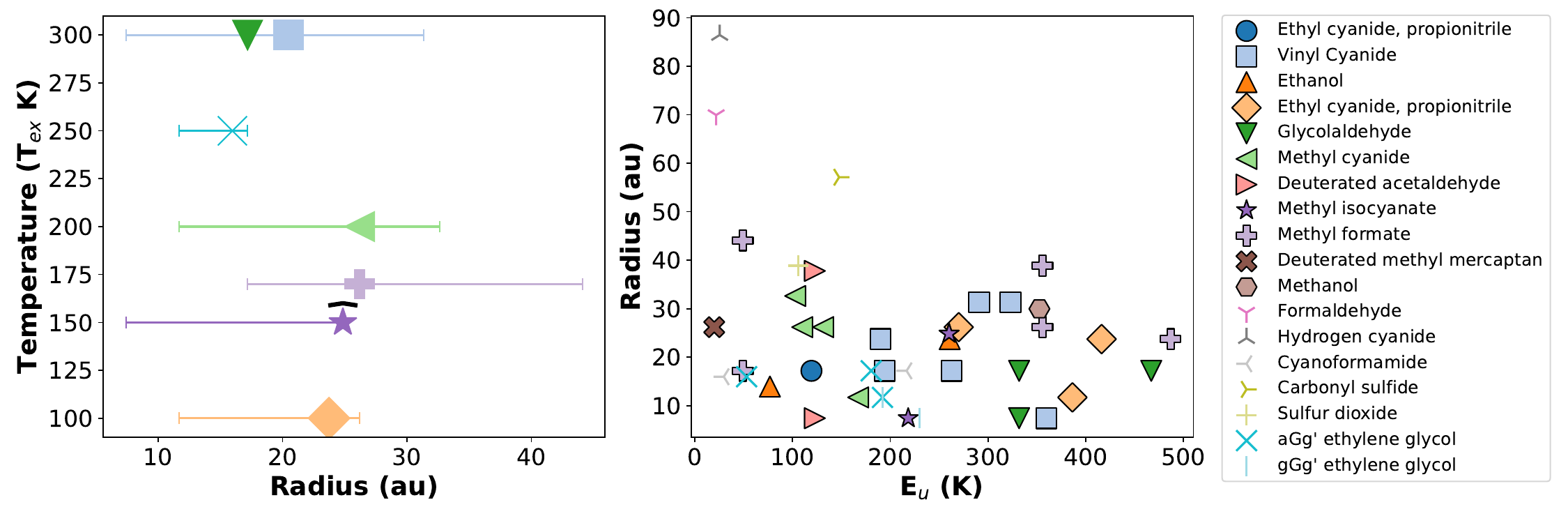}
\caption{Left panel: Deconvolved radii of the emission plotted as a function of excitation temperature for the confirmed detected molecules in IRAS4A2. The error bars represent the minimum and maximum radii observed for all molecular lines, while the markers indicate the median radius values. All molecules have at least one blended transition, except for methyl isocyanate, which is highlighted with a black hat on top of its marker to indicate the absence of blending. Right panel: The emission radius plotted against the upper state energy levels. Different colors and different figures represent emission from different molecules, as highlighted in the rightmost panel.} \label{fig: TempsAndDistribution}
\end{figure*}

The extended emission in Fig.\ref{fig: ExtendedEM} exhibits different morphologies, with some complexity. First, $\rm H_{2}CO$ is a well-established tracer of the gas disk in YSOs \citep{2020ApJ...901..166V,2021A&A...655A..65T}. For $\rm CH_{3}OCHO + SO$, the emission would appear to originate from the same region as $\rm H_{2}CO$, as they trace the same spatial extent,  but it could also be tracing a potential disk wind \citep{2024A&A...686L..13D}. Meanwhile, $\rm OCS$ appears to be an effective tracer of small grains, especially following the continuum emission bridge previously observed between IRAS4A1 and IRAS4A2 (as shown by the contours in Fig.~\ref{fig: ExtendedEM} and in \citet{2024A&A...681A..82G}). For both $\rm OCS$ and $\rm SO_{2,v=0}$, it has been suggested that their extended emission in IRAS4A could originate from interactions between the outflow and the surrounding material \citep{2020A&A...637A..63T}. The integrated emission maps further support the idea that both the grains and $\rm OCS$ are influenced by the S-shaped morphology of the IRAS4A2 outflow. A similar trend is observed for $\rm HCN$, although in this case, its emission more clearly traces the shape of the disk continuum in the inner regions, suggesting it may be simultaneously tracing both the outflow and the gas component of the disk. 

On the other hand, $\rm H_{2}CO$ and $\rm SO_{2}$ do not fully align with shape of the continuum emission. They are more compact and do not reach the outflow. This difference may arise from the fact that $\rm OCS$ transitions to the gas phase at lower temperatures than $\rm H_{2}CO$ and $\rm SO_{2}$. Moreover, smaller dust grains, less affected by radial drift, tend to remain distributed over more extended colder regions, making $\rm OCS$ a suitable tracer of regions where only smaller dust grains are present. In contrast, the other two lines require higher temperatures, needing to be closer to the star to reach their sublimation temperatures, and therefore trace more evolved structures. Given this difference, one might expect that $\rm OCS$ and $\rm HCN$ trace a region with small dust grains, while $\rm H_{2}CO$ and $\rm SO_{2}$ likely trace more inner structures, possibly those related to the gas disk. As shown in Fig.~\ref{fig:PV-diagramas}, the kinematics of $\rm H_{2}CO$ and $\rm OCS$ appear more similar compared to the other molecules, suggesting they may be tracing the same or closely related physical components. However, the moment 1 map of $\rm OCS$ differs in shape from those of the other molecules, particularly at larger offsets. In the inner regions, all molecules display a clearer velocity gradient, but the deviation observed in $\rm OCS$ at greater distances may indicate that its velocity map is influenced by additional components traced by this molecule. Additionally, it is important to note that the fit for $\rm H_{2}CO$ is not ideal, and much of the negative-velocity emission is not well reproduced. The interpretation for this line therefore remains subject to uncertainties. The main issues probably come from the poor S/N at the current resolution and the absorption features present in some spectral channels. While there's not much we can do about the absorption, future observations with higher S/N and improved spectral resolution would help to obtain a more reliable fit for this particular molecule.

$\rm OCS$, $\rm HCN$, and $\rm H_{2}CO$ emission exhibits significant absorption by the dust continuum opacity in the inner regions of the IRAS4A2 disk, with $\rm HCO^{+}$ being completely absorbed. It is also worth noting that some of this molecular emission may be subject to self-absorption due to infalling material along the line of sight, which could either contribute to or even account for the observed absorption features, rather than dust absorption alone. In contrast, $\rm SO_{2}$ shows little or no absorption compared to the other extended molecules, suggesting it may trace higher layers of the source, possibly the envelope. This behavior could indicate that $\rm HCN$ and $\rm H_{2}CO$ emission is likely tracing a more middle region, while $\rm HCO^+$ emission, due to its strong absorption, may trace regions closer to the mid-plane of the source, where dust absorption from the shrinking background might still be significant. This stratification of molecular emission was also observed in HH212 by \citet{2022ApJ...937...10L}, in IRAS4A2 by \citet{2024A&A...686L..13D}, and is consistent with findings in HD 163296, where emission from the same molecules were found to trace a similar vertical order \citep{2023A&A...669A.126P}. 

In summary, our analysis highlights clear differences in the behavior of these molecules. For example, $\rm H_{2}CO$ and $\rm SO_{2}$ show more compact emission that does not extend into the outflow. $\rm H_{2}CO$ and $\rm OCS$ share a similar P–V diagram model fit and exhibit aligned masses. Moreover, all extended molecules ($\rm OCS$, $\rm HCN$, $\rm H_{2}CO$, $\rm HCO^{+}$, and $\rm SO_{2}$) display distinct absorption patterns, and both $\rm HCN$ and $\rm OCS$ moment 1 maps suggest the presence of two separate components. Together, these differences, and in some cases, shared behaviors, point to the possibility that the molecules are tracing multiple physical components simultaneously, even on very small scales (<20 au), with variations in vertical distribution across radii and emission extent. However, further observations are required to confirm which physical components are being traced and to disentangle possible chemical and structural stratification within IRAS4A2.

\subsubsection{Compact emission and the origin of the dust absorption}

As previously discussed, the compact COMs in Figure~\ref{fig: CompactEM} also reveal different morphologies. Some molecules, such as $\rm C_{2}H_{3}CN$ and $\rm CH_{3}^{18}OH$, exhibit extremely compact emission within radii smaller than 20 au, well below the beam size (0.065"). This indicates that their emission is still very unresolved, and, combined with the low signal-to-noise, makes accurate size estimation particularly challenging. 

Despite the uncertainties associated with signal-to-noise and the effects of absorption, the Keplerian model derived from $\rm H_{2}CO$ provides a reasonable fit to the $\rm CH_{3}CDO$ and $\rm CH_{3}OCHO$ emission shown in Figure~\ref{fig: CompactEM}. Moreover, the $\rm CH_{3}OCHO$ moment 1 map reveals a clearer velocity gradient perpendicular to the outflow axis, consistent with the rotation seen in the flattened envelope. Combined with the fact that both emissions are not aligned with the outflow direction, this suggests that these molecules may likely serve as reliable tracers of the disk itself. Further investigation of these two molecules may provide valuable information about the embedded disk and the early conditions enabling planet formation.

Some of these compact regions appear to trace the innermost part of the IRAS4A2 system, as shown in Figure~\ref{fig: TempsAndDistribution}, especially given that emission from a couple of molecules remains unresolved. Moreover, low- and high-energy COM transitions probe a wide range of regions across the system. The most compact emission among these low upper state energy level lines could likely originate from colder regions in higher layers of the disk if the temperatures are lower there than in the mid-plane \citep{2021MNRAS.508.2583Z}, and if they are effectively shielded from the protostar’s radiation. Further supporting this interpretation is the observation that some compact and extended molecular emission still exhibits absorption features around IRAS4A2, whereas many others do not. This contrast implies that the emitting and sublimation regions of certain molecules may lie in higher and warmer layers of the disk structure, where the dust optically thick absorption appears to be now limited only to the innermost regions with lower scale height.

In Class 0/I objects, the molecular emission often seems to be absorbed over large spatial scales when having an optically thick dust background at sub-millimeter wavelengths \citep{2018ApJ...868...39G,2019A&A...632A...5G}. These absorption features have been observed and studied in other star-forming regions as well. For example, \citet{2019NatAs...3..314L} suggests that the absorption and reduction of line emission may be due to additional heating sources, which could be altering the temperature structure in the disk. This view comes from observations of systems experiencing accretion outbursts, similar to what has been observed in V883 Ori \citep{2024MNRAS.527.9655A,2024AJ....167...66Y}. In the context of young stellar objects (YSOs), simulations by \citet{2014ApJ...786...35B} suggest that viscous heating could significantly impact the disk temperature beyond 20 au, affecting both the mid-plane and higher layers of the disk. Moreover, various hot spots have been identified in other protostellar sources, such as IRAS 16293-2422 A  \citep{2022ApJ...941L..23M}. While the exact origin of these hot spots remains unclear, accretion shocks, particularly those associated with infalling material along filamentary structures (streamers) from the extended envelope, are a primary candidate for generating such localized heating. As highlighted in \citet{2022ApJ...941L..23M}, analyzing and identifying these diverse heating mechanisms is crucial for accurately modeling Class 0, and even Class I, protoplanetary disks as these systems have shown signs of late-stage accretion shocks \citep[e.g.,][]{2022A&A...658A.104G}

While additional heating processes may help explain some of the absorption features observed in IRAS4A2, another important factor to consider is the effect of continuum over-subtraction. It is crucial to distinguish between true absorption, which occurs when lower-excitation material lies in front of or above a warmer, dust continuum \citep{2019ApJ...885...98S}, and over-subtraction, an effect that arises when the optically thick continuum and the overlying optically thick gas have similar excitation temperatures, leading to absorption in the line profiles after subtraction of the continuum \citep[see][]{2024A&A...686A.201N}. In the case of IRAS4A2, although several portions of the spectrum exhibit negative features, some due to true absorption and some to noise, many are most likely the result of over-subtraction. Regardless of the underlying mechanism, this behavior is particularly clearly observed in the inner regions of IRAS4A2, very close to the central star. This behavior could suggest an evolutionary track, as previously mentioned, where strong absorption is now confined to the innermost regions for some molecules, in contrast to IRAS4A1 and some other regions, where much of its emission is still absorbed. This difference indicates that the optically thick dust background has either dissipated in the outer regions, has become optically thin in most areas, or has shrunk and moved closer to the star and the mid-plane (as large dust grains are affected by settling and rapid radial drift). While these findings are intriguing, the absorbed emission may obscure critical details, preventing us from seeing the full picture with current observational capabilities. Furthermore, drawing conclusions about high-temperature and more compact molecular tracers remains challenging due to the limitations of both the CASSIS model and our spectral data (see Section~\ref{sec:Results}), which explains the lack of a clear correlation between the excitation temperature and the radius of the emission.

\subsection{Accretion Shocks}

$\rm OCS$ and $\rm SO_{2}$ emission exhibits strong emission peaks that are offset from the peak of the continuum emission (see Fig.~\ref{fig: ExtendedEM}). This observed emission pattern could be linked to shocks, as discussed in \citet{2021A&A...653A.159V}, where sulfur-bearing molecules serve as tracers for material flowing from the envelope into the disk, likely originating from filaments formed by what are commonly referred to as streamers. The formation of $\rm SO$ and $\rm SO_{2}$ typically occurs towards the end of a shock, suggesting that their emission is tracing such processes.

The origin of shocks in protostellar systems could be attributed to various mechanisms, including outflows \citep{2012A&A...541A..39P}, accretion shocks \citep{2014Natur.507...78S, 2022A&A...658A.104G}, or disk winds \citep{2017A&A...607L...6T}. Notably, \citet{2022A&A...667A..20A} identified $\rm SO_{2}$ as the most reliable tracer of accretion shocks, while \citet{2022ApJ...941L..23M} suggested that such accretion shocks could be responsible for generating hot dust spots. Furthermore, $\rm OCS$ and $\rm SO_{2}$ emission appears too extended to be solely associated with the envelope \citep{2021A&A...646A..72H}. 

In IRAS4A2, shocks are a plausible mechanism behind the observed emission of $\rm OCS$ and $\rm SO_{2}$. Identifying a single, dominant origin for these shocks remains challenging, however, as they can arise in various contexts within the system. A key distinction lies between accretion shocks, which occur when infalling material decelerates upon impacting the disk surface, generating localized heating, and outflow-induced shocks, which result from the interaction between high-velocity outflows and the surrounding envelope, often producing more spatially extended emission with broader line profiles. Given the observations shown in Fig.~\ref{fig: ExtendedEM}, we propose that the compact, localized intensity peaks seen in $\rm OCS$ and $\rm SO_{2}$ integrated emission maps are most consistent with accretion shocks. Meanwhile, the more extended and broader emission features may instead be associated with outflow-driven shocks \citep{2020A&A...637A..63T}, reinforcing the idea that emission from these molecules could be tracing multiple components of the system.

It is difficult to determine from the P-V diagrams whether or not $\rm SO_{2}$ exhibits the symmetric Keplerian signature expected from a disk wind \citep{2017A&A...607L...6T}. If the absorption feature in the $\rm OCS$ P-V diagram is ignored, however, a pattern emerges, resembling the expected kinematics (broad and slower than the jet) of a rotating disk wind. This could explain why $\rm OCS$ appears to trace the smallest grains and suggests that it is simultaneously probing multiple components, such as accretion shocks and the disk wind. Nonetheless, determining the precise origin of these emissions remains challenging, and it is still possible that these lines trace a different component than assumed here.

\begin{figure}[t!]
\centering

\includegraphics[trim=0.0cm 0cm 0.0cm 0.0cm, clip=true,width=1\columnwidth]{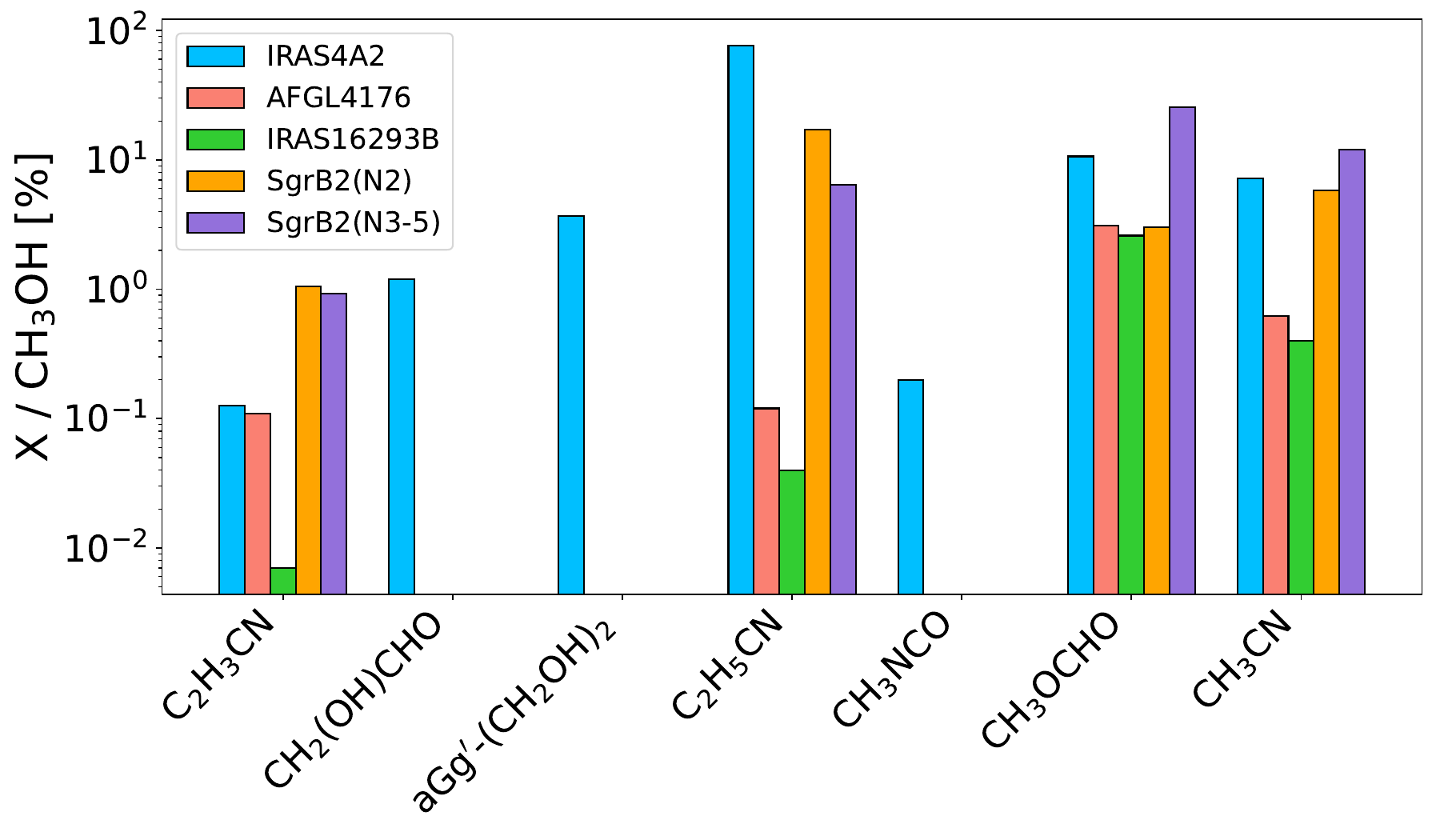}
\caption{Normalized molecular abundance ratios relative to methanol, for the species detected in this work. The values are compared to those observed in both high-mass star-forming regions (AFGL4176; \citealt{2019A&A...628A...2B} and SgrB2; \citealt{2017A&A...601A..49B}) and a low-mass star-forming region (IRAS16293B; \citealt{2016A&A...595A.117J}).} \label{fig: Methanolratios}
\end{figure}

\subsection{High mass star forming regions vs IRAS4A2}

To understand the origin of hot corinos, and in particular the one associated with IRAS4A2, it is valuable to compare their chemical composition with those of hot cores of other star-forming regions and identify both similarities and differences. Hot corinos have been proposed as low-mass analogs of the hot cores observed in high-mass star-forming regions, sharing comparable chemical signatures despite differences in mass and environmental conditions. In Fig.~\ref{fig: Methanolratios}, we compare the molecular abundances relative to methanol for the species detected in IRAS4A2 with those reported in both high-mass \citep{2019A&A...628A...2B,2017A&A...601A..49B} and low-mass star-forming regions \citep{2016A&A...595A.117J}. The column density of methanol was adopted from \citet{2025A&A...695A..78F}, using the value that does not account for dust absorption effects for consistency with this work. When necessary, isotopologue abundances were adjusted to reflect the main species using solar system isotopic ratios from \citet{1994ARA&A..32..191W}.

From Fig.~\ref{fig: Methanolratios}, IRAS4A2 appears to exhibit molecular abundances similar to those of high-mass star-forming regions such as AFGL4176 and Sgr B2, at least for the molecules where direct comparisons are possible. In the case of $\rm C_{2}H_{5}{}CN$, where we predict a high molecular abundance, we suspect the CASSIS fitting may be unreliable because only one unblended line is clearly detected in our spectra. The remaining lines are either heavily blended or faint, introducing significant uncertainty in the total molecular abundance of this molecule.

When comparing the molecular inventory of IRAS4A2 with that of the high-mass star-forming region AFGL 4176 in \citet{2019A&A...628A...2B}, we find that both regions share many molecular species, though AFGL 4176 exhibits a greater number of transitions and lines, probably due to its higher mass. Interestingly, the only molecules not identified in AFGL 4176 are  $\rm NCC(O)NH_2$ and $\rm CH_3NCO$. $\rm CH_3NCO$, however, has been detected in other high-mass star-forming regions, including Orion KL and Sgr B2(N) \citep{2016A&A...587L...4C, 2017A&A...601A..49B}, and a tentative detection of $\rm NCC(O)NH_2$ has recently been reported in Sgr B2(N) \citep{2024PASJ...76...54L}. This absence is not entirely surprising, as previous studies have noted that AFGL 4176 is more comparable to the low-mass protobinary system IRAS 16293 \citep{2016A&A...595A.117J} than to other high-mass sources like Sgr B2(N) and Orion KL. In contrast, IRAS4A2 appears more aligned with the latter, supporting the idea that hot corinos may be low-mass analogs of the hot cores in high-mass star-forming regions, sharing similar complex chemistry but on a smaller scale.

Regarding the spatial distribution of these molecules, we observe several similarities with AFGL 4176 \citet{2019A&A...628A...2B}. First, emission from transitions with lower upper-state energies appears to be more spatially extended than that from higher upper-state energy transitions. Additionally, emission from sulfur-bearing molecules is among the most extended compared to emission from other molecules, which aligns with detections in AFGL 4176 and our spatially resolved observations in IRAS4A2. Finally, we also observe a reverse trend with $\rm C_2H_3CN$, like in AFGL 4176, with some higher upper-state energy transitions corresponding to larger radii. The reason for this reversal is still unclear and needs more investigation.

\section{Summary and conclusions}
We have studied the hot corino of IRAS4A2 at high resolution using ALMA, analyzing different spatial scales, from large-scale outflow-like emission down to a small warm inner envelope or disk ($\sim$20 au) emission. Our findings indicate that:
\begin{itemize}
    \item We confirm the detection of four complex organic molecules around IRAS4A2: $\rm C_{2}H_{3}CN$, $\rm CH_{2}(OH)CHO$, $\rm CH_{3}OCHO$, and $\rm CH_{3}C^{15}N$, along with tentative detections of $\rm aGa'$-glycol, $\rm C_{2}H_{5}^{13}CN$, and $\rm CH_{3}NCO$. This is the first time $\rm C_{2}H_{3}CN$ and $\rm aGg'$-glycol are detected towards IRAS4A2.
    \item Molecular emission appears to trace distinct regions within the IRAS4A2 system; however, determining their exact origins remains challenging with the current data. Several species with small spatial extents and high excitation temperatures are likely associated with the innermost regions of IRAS4A2. In contrast, molecules such as $\rm CH_{3}CDO$ and $\rm CH_{3}OCHO$ may serve as effective tracers of the disk structure. Further studies are necessary to determine the potential of these molecules for tracing the disk. If confirmed, however, these molecules could serve as key tracers for studying the disk during its earliest stages, when planet formation may have already begun.
    \item  Given the inconsistencies between absorption features, kinematic signatures (e.g., P–V diagrams), and other observed properties, it is likely that extended emission is tracing multiple components simultaneously. These components include the outflow, envelope, disk, warm inner envelope, and possibly accretion shocks, outflow shocks and the disk wind, reflecting the complex and dynamic environment of this young protostellar system.
    \item The emission peaks observed in the $\rm OCS$ and $\rm SO_{2}$ moment 0 maps are more consistent with signatures typically associated with shocks, particularly accretion shocks, which may be responsible for heating the dust in IRAS4A2. This hot dust could explain the prominent absorption features detected across the innermost regions of the spectra. In contrast, the more extended emission coming from the same molecules is better aligned with characteristics of outflow-driven shocks, suggesting that multiple shock mechanisms are contributing to the observed molecular emission.
    \item A comparison with high-mass star-forming regions such as AFGL 4176, Sgr B2(N), and Orion KL shows that these sources share mostly the same molecular species. This commonality suggests that hot corinos do not exhibit significant differences in molecular composition regardless of mass or environment. Furthermore, the spatial distribution of these molecules appears to be similar across different sources, but more studies at high resolution are needed to confirm this.
    \item We resolved the large-scale ($\sim$5300 au) outflows of $\rm HCN$, $\rm HCO^{+}$, and $\rm H_{2}CO$, distinguishing the outflows from IRAS4A1 and IRAS4A2. We confirm the previously reported S-shaped morphology and the inverted velocity gradient observed in this region, between the outflow emission and other components of IRAS4A2.

\end{itemize}

\begin{acknowledgements}
 We thank J. Alejandro L\'opez-V\'azquez, Annaëlle Maury, and Marta de Simone for their helpful discussions and insights. We acknowledge assistance from Allegro, the European ALMA Regional Centre node in the Netherlands. This paper makes use of the following ALMA data: ADS/JAO.ALMA$\#$2018.1.00510.S. ALMA is a partnership of ESO (representing its member states), NSF (USA), and NINS (Japan), together with NRC (Canada), MOST and ASIAA (Taiwan), and KASI (Republic of Korea), in cooperation with the Republic of Chile. The Joint ALMA Observatory is operated by ESO, AUI/NRAO and NAOJ. P.N. acknowledges support from the ESO and IAU Gruber Foundation Fellowships.

\end{acknowledgements}

\bibliographystyle{aa}
\bibliography{bibliography.bib}

\onecolumn
\begin{appendix}
\section{Spectral analysis and CASSIS modeling of the IRAS4A2 region}

\begin{figure*}[ht!]
\centering

\includegraphics[trim=0 0cm 0 0, clip=true,width=0.75\textwidth]{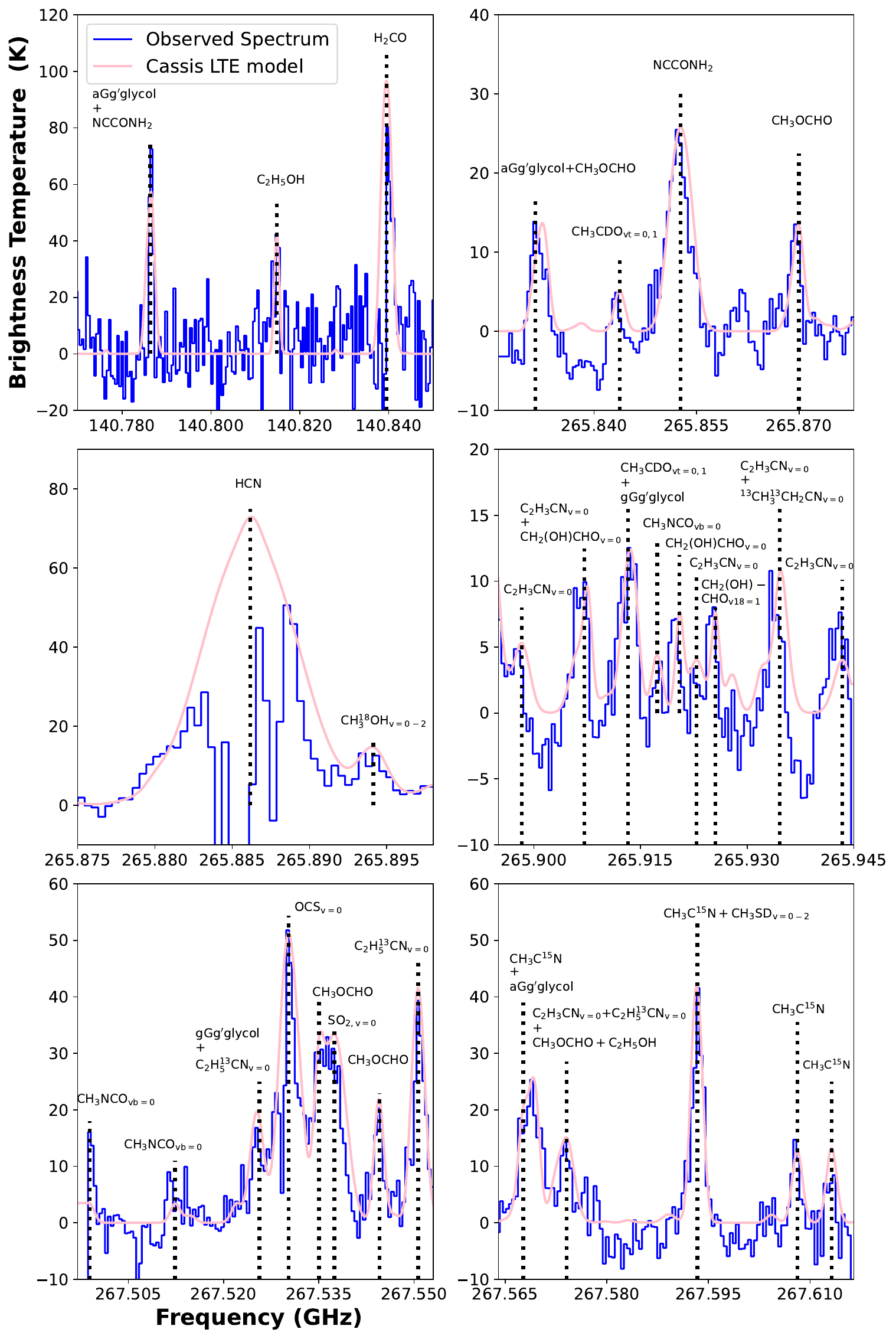}
\caption{CASSIS models of the spectral lines in IRAS4A2 (pink) are compared with the observed spectra (blue) at frequencies where line emission is detected. The molecules are labeled in black } \label{fig: Cassis Model}
\end{figure*}
\clearpage

\section{IRAS4A2 Molecules in spectra}
\label{appendix:a}
\begin{table*}[h]
 \centering
 \setlength\tabcolsep{5.pt}
\renewcommand{\arraystretch}{1.3}
  \captionsetup{justification=centering}
    \caption{Line characteristics}
\resizebox{1\textwidth}{!}{%
\begin{tabular}{ccccccc}
 \hline
    Molecule & Name & Transition & Rest Frequency (GHz) & A$_{ij}$ (s$^{-1}$) & E$_{u}$ (K) & g$_{u}$ \\
 \hline
 $\rm HCN$ & Hydrogen cyanide & (3-2) & 265.88618 & 8.42e-4 & 25.52 & 21 \\
 $\rm C_2H_3CN_{v=0}$ & Vinyl Cyanide & (28 6 23-27 6 22) & 265.9347953 & 1.49e-3 & 262.77 & 171 \\
 $\rm C_2H_3CN_{v=0}$ & Vinyl Cyanide & (28 8 21-27 8 20)(28 8 20-27 8 19) & 265.9062054 & 1.44e-3 & 323.06 & 171 \\
 $\rm C_2H_3CN_{v=0}$ & Vinyl Cyanide & (28 9 19-27 9 18)(28 9 20-27 9 19) & 265.943385 & 1.40e-3 & 359.59 & 171 \\
 $\rm C_2H_3CN_{v=0}$ & Vinyl Cyanide & (29 1 29-28 1 28) & 265.9228854 & 1.56e-3 & 194.36 & 177 \\
 $\rm C_2H_3CN_{v=0}$ & Vinyl Cyanide & (28 7 21-27 7 20) & 265.8982758 & 1.47e-3 & 290.78 & 171 \\
 $\rm C_2H_3CN_{v=0}$ & Vinyl Cyanide & (28 1 27-27 1 26) & 267.5746326 & 1.59e-3 & 189.95 & 171 \\
 $\rm CH_3CDO_{vt=0,1}$ & Deuterated acetaldehyde & (14 4 10 0-13 4 9 0) & 265.91423 & 6.12e-4 & 122.84 & 29 \\
 $\rm CH_3CDO_{vt=0,1}$ & Deuterated acetaldehyde & (14 4 10 2-13 4 9 2) & 265.843748 & 6.05e-4 & 122.83 & 29 \\
 $\rm CH_2(OH)CHO_{v=0}$ & Glycolaldehyde & (34 3 31 0-34 2 32 0) & 265.9076199 & 1.51e-4 & 331.78 & 69 \\
 $\rm CH_2(OH)CHO_{v=0}$ & Glycolaldehyde & (34 4 31 0-34 3 32 0) & 265.9204834 & 1.51e-4 & 331.78 & 69 \\
 $\rm CH_2(OH)CHO_{v18=1}$ & Glycolaldehyde & (18 11 7 1-18 10 8 1) & 265.9255426 & 2.20e-4 & 466.95 & 37 \\
 $\rm CH_3{}^{18}OH_{v=0-2}$ & Methanol & (16 3 14 0-16 2 15 0) & 265.894132 & 9.63e-5 & 352.57 & 132 \\
 $\rm NCC(O)NH_2$ & Cyanoformamide & (23 18 5-23 17 6)(23 18 6-23 17 7) & 265.852665 & 3.14e-4 & 216.70 & 47 \\
 $\rm NCC(O)NH_2$ & Cyanoformamide & (11 4 7-10 3 8) & 140.7863341 & 4.88e-5 & 29.71 & 23 \\
 $\rm gGg'-(CH_{2}OH)_{2}$ & gGg' ethylene glycol & (26 11 16 1-25 11 15 0) & 265.9131258 & 1.42e-4 & 230.25 & 371 \\
 $\rm gGg'-(CH_{2}OH)_{2}$ & gGg' ethylene glycol & (27 3 24 1-26 4 23 1) & 267.5258901 & 2.10e-4 & 191.86 & 385 \\
 $\rm aGg'-(CH_{2}OH)_{2}$ & aGg' ethylene glycol & (26 4 23 1-25 3 22 1) & 265.8326715 & 2.06e-4 & 180.49 & 371 \\
 $\rm aGg'-(CH_{2}OH)_{2}$ & aGg' ethylene glycol & (26 6 20 0-25 6 19 1) & 267.5692918 & 4.29e-4 & 192.05 & 371 \\
 $\rm aGg'-(CH_{2}OH)_{2}$ & aGg' ethylene glycol & (13 4 10 1-12 4 9 0) & 140.7865974 & 6.02e-5 & 53.05 & 243 \\
 $\rm {}^{13}CH_3{}^{13}CH_2CN$ & Ethyl cyanide, propionitrile & (12 9 4-13 8 5)(12 9 3-13 8 6) & 265.9346645 & 5.15e-6 & 119.36 & 25 \\
 $\rm C_2H_5{}^{13}CN_{v=0}$ & Ethyl cyanide, propionitrile & (30 8 23-29 8 22) & 267.5506331 & 1.48e-3 & 270.05 & 61 \\
 $\rm C_2H_5{}^{13}CN_{v=0}$&Ethyl cyanide, propionitrile& (30 14 17-29 14 16)(30 14 16-29 14 15)  & 267.5740406  &1.25e-3 &416.15&61\\
$\rm C_2H_5{}^{13}CN_{v=0}$&Ethyl cyanide, propionitrile& (30 13 18-29 13 17)(30 13 17-29 13 16) & 267.5248859 &1.30e-3 &386.33&61\\
$\rm CH_3NCO_{vb=0}$&Methyl isocyanate& (31 0 0 1-30 0 0 1) & 265.917341  &9.14e-4 & 218.36&63 \\
$\rm CH_3NCO_{vb=0}$&Methyl isocyanate& (31 3 28 0-30 3 27 0) &267.512332  &9.12e-4 & 260.16&63\\
$\rm CH_3NCO_{vb=0}$&Methyl isocyanate& (31 3 29 0-30 3 28 0)& 267.498881  &9.12e-4 &260.15&63\\
$\rm CH_3OCHO$&Methyl formate& (7 7 0 0-6 6 1 0)(7 7 1 0-6 6 0 0) & 265.869967 &4.35e-5 & 49.05&30\\
$\rm CH_3OCHO$&Methyl formate& (7 7 0 2-6 6 0 2)& 265.831401 &4.35e-5 & 49.07&30 \\
$\rm CH_3OCHO$&Methyl formate& (24 1 23 3-23 1 22 3)& 267.544467 &2.85e-4 & 355.60&98\\
$\rm CH_3OCHO$&Methyl formate& (24 2 23 3-23 2 22 3)& 267.535013 &2.85e-4 & 355.60&98\\
$\rm CH_3OCHO$&Methyl formate& (22 15 7 5-21 15 6 5)&267.572288 &1.55e-4 & 486.69&90\\
$\rm CH_3C{}^{15}N$&Methyl cyanide& (15 1-14 1) & 267.608 &1.66e-3 & 109.91&62\\
$\rm CH_3C{}^{15}N$&Methyl cyanide& (15 3-14 3)&267.567 &1.60e-3 & 167.16&124\\
$\rm CH_3C{}^{15}N$&Methyl cyanide&(15 2-14 2) &267.593&1.63e-3 & 131.38&62\\
$\rm CH_3C{}^{15}N$&Methyl cyanide&(15 0-14 0)&267.613 &1.66e-3 & 102.76&62\\
$\rm HCO^{+}_{v=0,1,2}$&Formyl ion&(3 0 0-2 0 0)&267.557526 &1.45e-3 & 25.68&7\\
$\rm CH_3SD_{v=0-2}$&Deuterated methyl mercaptan &(3 2 1 1-2 1 1 1) &267.599 &2.56e-5 & 19.95&7\\
$\rm SO_{2}~_{v=0}$&Sulfur dioxide&(13 3 11-13 2 12)&267.5374512 &1.51e-4 & 105.82&27\\
$\rm OCS_{v=0}$&Carbonyl sulfide&(22-21) &267.530219 &5.57e-5 & 147.67&45\\
$\rm H_2CO$&Formaldehyde&(2 1 2-1 1 1)&140.839502 &5.30e-5 & 21.92&15\\
$\rm C_2H_5OH$&Ethanol&(12 3 10 2-12 2 11 2)&140.81481 &1.67e-5 &76.93&25\\
$\rm C_2H_5OH$&Ethanol&(15 9 6 1-16 8 8 0) &267.5732224 &7.63e-6 &260.82&31 \\
 \hline
\end{tabular}
}
\label{tab:IRAS4A2 Molecules First Table}
\end{table*}

\begin{table*}[h]
 \centering
 \setlength\tabcolsep{3.pt}
 \renewcommand{\arraystretch}{1.5}
 \captionsetup{justification=centering}
 \caption{Physical properties and observed sizes}
 \resizebox{1\textwidth}{!}{%
\begin{tabular}{ccccccc}
 \hline
 Molecule & Name & Transition & FWHM [Km/s] & Observed image sizes (au) & T$_{ex}$ [K] & N$_{Sp}$ [cm$^{-2}$] \\
 \hline
 $\rm HCN$ & Hydrogen cyanide & (3-2) & 7.5 & 86 & - & - \\
 \hline
 $\rm C_2H_3CN_{v=0}$ & Vinyl Cyanide & (28 6 23-27 6 22) & 3 & 17 & 300 $\pm$ 4 & 1.77e15 $\pm$ 0.01e15 \\
 $\rm C_2H_3CN_{v=0}$ & Vinyl Cyanide & (28 8 21-27 8 20)(28 8 20-27 8 19) & 3 & 31 &  &  \\
 $\rm C_2H_3CN_{v=0}$ & Vinyl Cyanide & (28 9 19-27 9 18)(28 9 20-27 9 19) & 3 & 7 &  & \\
 $\rm C_2H_3CN_{v=0}$ & Vinyl Cyanide & (29 1 29-28 1 28) & 3 & 17 & &  \\
 $\rm C_2H_3CN_{v=0}$ & Vinyl Cyanide & (28 7 21-27 7 20) & 3 & 31 &  & \\
 $\rm C_2H_3CN_{v=0}$ & Vinyl Cyanide & (28 1 27-27 1 26) & 3 & 23 &  & \\
\hline
 $\rm CH_3CDO_{vt=0,1}$ & Deuterated acetaldehyde & (14 4 10 0-13 4 9 0) & 2 & 7 & - & - \\
 $\rm CH_3CDO_{vt=0,1}$ & Deuterated acetaldehyde & (14 4 10 2-13 4 9 2) & 2 & 38 & - & - \\
  \hline
 $\rm CH_2(OH)CHO_{v=0}$ & Glycolaldehyde & (34 3 31 0-34 2 32 0) & 1.5 & 7 & 300 $\pm$ 3 & 1.68e16 $\pm$ 0.1e16 \\
 $\rm CH_2(OH)CHO_{v=0}$ & Glycolaldehyde & (34 4 31 0-34 3 32 0) & 1.5 & 17 &  &  \\
 $\rm CH_2(OH)CHO_{v18=1}$ & Glycolaldehyde & (18 11 7 1-18 10 8 1) & 1.5 & 17 &  &  \\
  \hline
 $\rm CH_3{}^{18}OH_{v=0-2}$ & Methanol & (16 3 14 0-16 2 15 0) & 2.5 & 30 & - & - \\
  \hline
 $\rm NCC(O)NH_2$ & Cyanoformamide & (23 18 5-23 17 6)(23 18 6-23 17 7) & 4 & 17 & - & - \\
 $\rm NCC(O)NH_2$ & Cyanoformamide & (11 4 7-10 3 8) & 4 & 16 & - & - \\
  \hline
 $\rm gGg'-(CH_{2}OH)_{2}$ & gGg' ethylene glycol & (26 11 16 1-25 11 15 0) & 2.5 & 7 & - & - \\
 $\rm gGg'-(CH_{2}OH)_{2}$ & gGg' ethylene glycol & (27 3 24 1-26 4 23 1) & 2.5 & 11 & - & - \\
  \hline
 $\rm aGg'-(CH_{2}OH)_{2}$ & aGg' ethylene glycol & (26 4 23 1-25 3 22 1) & 2 & 17 & 250 $\pm$ 3 & 5.13e16 $\pm$ 0.02e16 \\
 $\rm aGg'-(CH_{2}OH)_{2}$ & aGg' ethylene glycol & (26 6 20 0-25 6 19 1) & 2 & 11 & & \\
 $\rm aGg'-(CH_{2}OH)_{2}$ & aGg' ethylene glycol & (13 4 10 1-12 4 9 0) & 2 & 16 &  &  \\
  \hline
 $\rm {}^{13}CH_3{}^{13}CH_2CN$ & Ethyl cyanide, propionitrile & (12 9 4-13 8 5)(12 9 3-13 8 6) & 2 & 17 & - & - \\
  \hline
 $\rm C_2H_5{}^{13}CN_{v=0}$ & Ethyl cyanide, propionitrile & (30 8 23-29 8 22) & 2.5 & 26 & 100 $\pm$ 0.3 & 1.21e16 $\pm$ 0.37e16 \\
 $\rm C_2H_5{}^{13}CN_{v=0}$ & Ethyl cyanide, propionitrile & (30 14 17-29 14 16)(30 14 16-29 14 15) & 2.5 & 23 &  &  \\
 $\rm C_2H_5{}^{13}CN_{v=0}$ & Ethyl cyanide, propionitrile & (30 13 18-29 13 17)(30 13 17-29 13 16) & 2.5 & 11 &  &  \\
  \hline
 $\rm CH_3NCO_{vb=0}$ & Methyl isocyanate & (31 0 0 1-30 0 0 1) & 2 & 7 & 150 $\pm$ 4 & 2.8e15 $\pm$ 0.05e15 \\
 $\rm CH_3NCO_{vb=0}$ & Methyl isocyanate & (31 3 28 0-30 3 27 0) & 2 & 24 &  &  \\
 $\rm CH_3NCO_{vb=0}$ & Methyl isocyanate & (31 3 29 0-30 3 28 0) & 2 & 24 &  &  \\
  \hline
 $\rm CH_3OCHO$ & Methyl formate & (7 7 0 0-6 6 1 0)(7 7 1 0-6 6 0 0) & 2 & 44 & 170 $\pm$ 3 & 1.49e17 $\pm$ 0.09e17 \\
 $\rm CH_3OCHO$ & Methyl formate & (7 7 0 2-6 6 0 2) & 2 & 17 &  &  \\
 $\rm CH_3OCHO$ & Methyl formate & (24 1 23 3-23 1 22 3) & 2 & 26 &  &  \\
 $\rm CH_3OCHO$ & Methyl formate & (24 2 23 3-23 2 22 3) & 2 & 39 &  &  \\
 $\rm CH_3OCHO$ & Methyl formate & (22 15 7 5-21 15 6 5) & 2 & 24 &  &  \\
  \hline
 $\rm CH_3C{}^{15}N$ & Methyl cyanide & (15 1-14 1) & 2 & 26 & 200 $\pm$ 6 & 3.73e14 $\pm$ 0.2e14 \\
 $\rm CH_3C{}^{15}N$ & Methyl cyanide & (15 3-14 3) & 2 & 11 &  &  \\
 $\rm CH_3C{}^{15}N$ & Methyl cyanide & (15 2-14 2) & 2 & 26 &  & \\
 $\rm CH_3C{}^{15}N$ & Methyl cyanide & (15 0-14 0) & 2 & 33 & &  \\
  \hline
 $\rm HCO^{+}_{v=0,1,2}$ & Formyl ion & (3 0 0-2 0 0) & - & - & - & - \\
  \hline
 $\rm CH_3SD_{v=0-2}$ & Deuterated methyl mercaptan & (3 2 1 1-2 1 1 1) & 2 & 26 & - & - \\
  \hline
 $\rm SO_{2}~_{v=0}$ & Sulfur dioxide & (13 3 11-13 2 12) & 4.5 & 39 & - & - \\
  \hline
 $\rm OCS_{v=0}$ & Carbonyl sulfide & (22-21) & 3.5 &  57 & - & - \\
  \hline
 $\rm H_2CO$ & Formaldehyde & (2 1 2-1 1 1) & 5 & 70 & - & - \\
  \hline
 $\rm C_2H_5OH$ & Ethanol & (12 3 10 2-12 2 11 2) & 2.5 & 14 & - & - \\
 $\rm C_2H_5OH$ & Ethanol & (15 9 6 1-16 8 8 0) & 2.5 &  24 & - & - \\
 \hline
\end{tabular}
}
\label{tab:IRAS4A2 Molecules Second Table}
\end{table*}
\clearpage

\section{IRAS4A2 Moment 0 maps}
\FloatBarrier
\begin{figure}[b!]
\centering

\includegraphics[trim=0.0cm 3cm 0.0cm 0.0cm, clip=true,width=0.9\columnwidth]{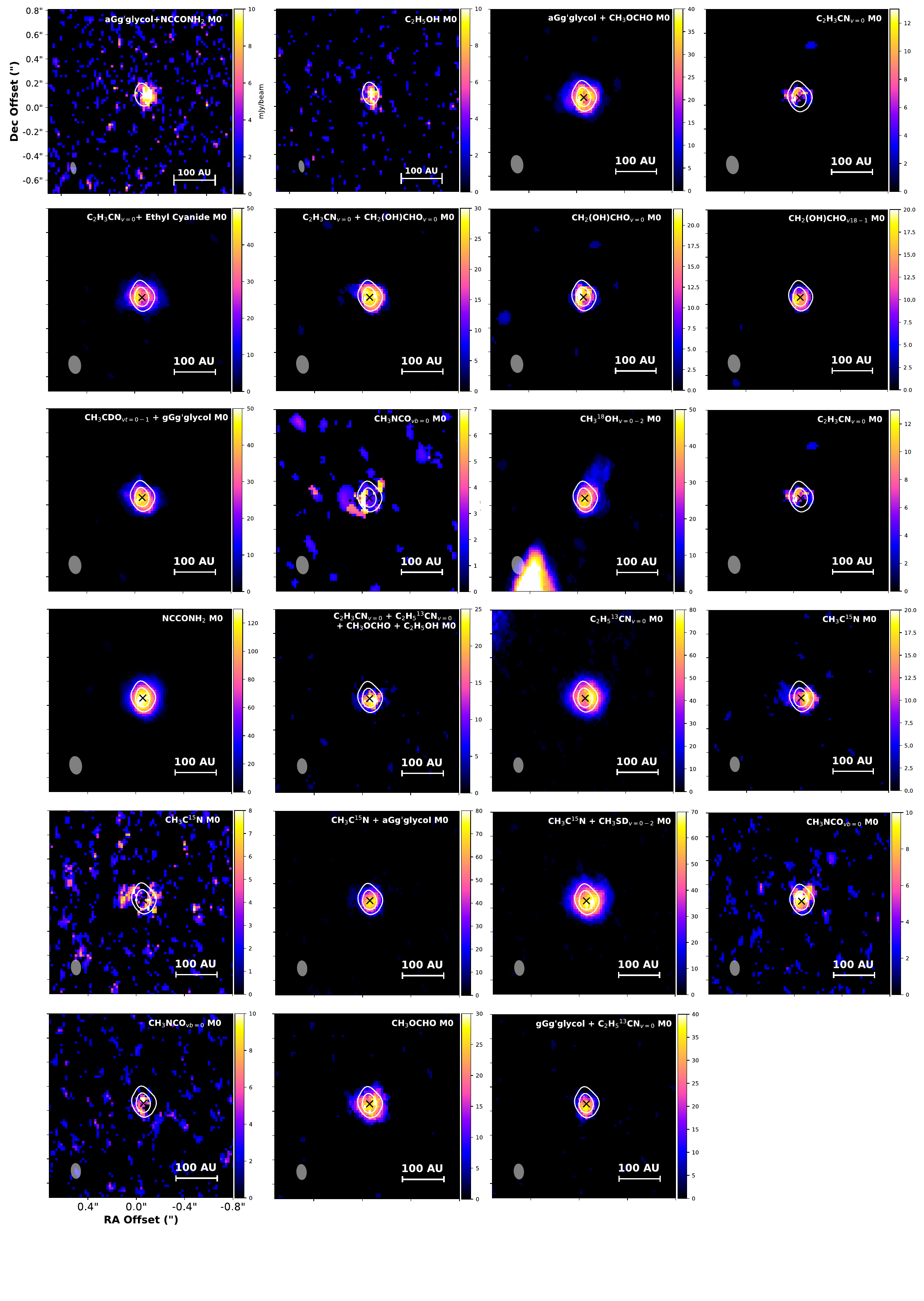}
\caption{Integrated emission (Moment 0) maps of all the remaining molecular lines identified in the ALMA Band 6 and Band 4 spectra of IRAS4A2.} \label{fig: Moment0maps}
\end{figure}
\label{appendix:b}
\end{appendix}

\end{document}